\magnification=\magstephalf
\baselineskip=12pt
\vsize=22.0truecm
\hsize=15.5truecm
\parindent=0.8truecm
\nopagenumbers
\parskip=0.2truecm
\def\vs{\vskip 0.2in}
\def\ts{\vskip 0.05in}

\def\n{\noindent}

\def\deg{^\circ}
\def\etal{{\it et al.\ }}
\def\Msun{${\rm M}_\odot\>$}
\font\bigbold=cmbx12
\font\quote=cmr9
\font\sm=cmr8

\centerline{{\bigbold The information content of gravitational wave}}

\centerline{{\bigbold harmonics in compact binary inspiral}} \vs \vs
\centerline {Ronald W. Hellings} \centerline {Department of Physics, Montana
State University, Bozeman MT 59715 } \vs \centerline {Thomas A. Moore}
\centerline {Department of Physics and Astronomy, Pomona College, Claremont CA
91109}

\vs

\n ABSTRACT:  The nonlinear aspect of gravitational wave generation that
produces power at harmonics of the orbital frequency, above the
fundamental quadrupole frequency, is examined to see what information
about the source is contained in these higher harmonics.  We use an order
(4/2) post-Newtonian expansion of the gravitational wave waveform of a
binary system to model the signal seen in a spaceborne gravitational wave
detector such as the proposed LISA detector.  Covariance studies are then
performed to determine the ultimate accuracy to be expected when the
parameters of the source are fit to the received signal.  We find three
areas where the higher harmonics contribute crucial information that
breaks degeneracies in the model and allows otherwise badly-correlated
parameters to be separated and determined.  First, we find that the
position of a coalescing massive black hole binary in an ecliptic plane
detector, such as OMEGA, is well-determined with the help of these
harmonics.  Second, we find that the individual masses of the stars in a
chirping neutron star binary can be separated because of the mass
dependence of the harmonic contributions to the wave.  Finally, we note
that supermassive black hole binaries, whose frequencies are too low to be
seen in the detector sensitivity window for long, may still have their
masses, distances, and positions determined since the information content
of the higher harmonics compensates for the information lost when the
orbit-induced modulation of the signal does not last long enough to be
apparent in the data.

\vfill \eject

\n {\bf I. Introduction}

As two compact stars in a tight binary system approach each other, the
nonlinear effects of relativistic gravity produce two effects in the waveform
of the gravitational wave emitted by the binary.  First, the loss of energy to
gravitational radiation tightens the binary, decreasing the gravitational wave
period and increasing its amplitude in a ``chirp''.  Second, the scattering of
the emitted gravitational waves off of the strong gravitational field around
the binary converts some of the energy to higher harmonics, modifying the
shape of the waves.  Several papers have been written [1,2] exploring the
sensitivity of the signals from these gravitational wave sources to the
parameters of the systems, but only one of these (Moore and Hellings [2],
hereafter referred to as MH) has taken the information content of the higher
harmonics into account.  It is the purpose of this paper to expand on MH by
particularly investigating the information content of the higher harmonics of
the gravitational wave waveform, identifying three places where these
harmonics break degeneracies in the solutions and allow otherwise
poorly-separated parameters to be determined.

The remainder of this paper is organized as follows.  In Section II, we
present the mathematical form of the gravitational waves from inspiraling
compact binaries, as seen in the detectors.  The detectors considered will
be the same as those in MH, namely an ecliptic-plane case, similar to the
proposed OMEGA mission [3], and a precessing-plane case, similar to the
proposed LISA mission [4]. In discussing the information content of the
various contributions to the received signal, we will need rather complete
expressions, so we will reproduce here the exposition as it is given in
MH. In sections III through V, we will discuss the three places where the
higher harmonics contribute to the determination of the binary parameters,
giving the results of covariance studies to predict the uncertainties in
the realistic parameter-estimation process.  For brevity, we will omit
many of the details of the covariance study procedure, referring the
reader instead to MH for more information.  Finally, in section VI, we
will draw a general conclusion from this study.

\vs \n {\bf II. The received gravitational wave waveform}

The gravitational wave from a binary system of non-spinning point masses in
quasi-circular orbit has been worked out by Blanchet \etal[5].  The waveform,
to order (4/2) in $v/c$, can be written in the frame of the detector as
$$\eqalign{
 h_{+,\times }(t)=&{2\tau c\eta\over5R_L}(1+z)\varepsilon ^2
 \left[H_{+,\times}^{(0)}+\varepsilon H_{+,\times }^{(1/2)}
 \right. \cr
& \left.+\varepsilon^2H_{+,\times}^{(2/2)}+\varepsilon^3H_{+,\times }^{(3/2)}
   +\varepsilon^4H_{+,\times}^{(4/2)} \right] ,
 } \eqno(1)$$
where $R_L$ is the luminosity distance to the source in an assumed flat
Friedman universe, $c$ is the speed of light, $\tau \equiv5G(m_1+m_2)/c^3$
is proportional to the total mass in units of time, $\eta \equiv m_1m_2/
(m_1+m_2)^2$ is the ratio of reduced mass to total mass, $\varepsilon
\equiv (\tau \omega_s/5)^{1/3}$ is a time-dependent expansion parameter
that is of order $v/c$ for the system, and $z$ is the cosmological
redshift.  The $\omega_s$ that appears in the definition of $\varepsilon$
is the time-dependent angular frequency of the binary orbit in its own
frame. The terms in the multipole expansion of the waveform are given by
$$
\eqalignno{ H_+^{(0)}&=-(1+\cos ^2i)\cos 2\phi _r &(2a) \cr H_\times
^{(0)}&=-2\cos i\;\sin 2\phi _r &(2b)\cr H_+^{(1/2)}&=-{\delta  \over
8}\sin i\left[ {(5+\cos ^2i)\cos \phi _r}\right. &(2c)\cr
&\quad\quad\quad\quad\left.{-9(1+\cos ^2i)\cos 3\phi _r}\right]\cr
H_\times ^{(1/2)}&=-{3\delta  \over 4}\cos i \sin i\left[ \sin \phi
_r-3\sin 3\phi _r\right] &(2d)\cr &{\rm etc.\, ,}} $$ where $i$ is the
angle of the source's orbital angular momentum vector relative to a unit
vector pointing from the source to the detector, $\delta \equiv
(m_1-m_2)/( m_1+m_2)$, and $\phi _r$ is the received phase as a function
of time.  The complete expressions for the higher-order $H$s are found in
Blanchet \etal [5].

To the level of approximation required, the phase received at time $t$ is the
Doppler-shifted version of the phase of the signal at the source
$$
\phi_r(t)=\phi(t_s)-{\Omega R\over c(1+z)}\sin\Theta I_0(t)+\phi_0 , \eqno
(3a)
$$
where
$$
I_0(t)\equiv\int_0^t\omega_s\sin(\Omega t-\Phi)dt . \eqno(3b)
$$
In Eqs. 3, $t_s=t/(1+z)$ is the time at the source, $\Omega$ and $R$ are
the angular frequency and radius of the detector's orbit around the sun,
$\Theta$ and $\Phi$ are the source's ecliptic coordinates in the sky
(oriented so that $\Phi=0$ corresponds to the direction of the detector
relative to the sun at $t=0$), and $\phi_0$ is the phase of the received
signal at $t=0$.

According to Blanchet \etal [5], the orbital phase is in turn given by
$$
\phi (t_s)={1 \over \eta }\left[ {F(t_s)-F(0)}\right] \eqno (4a)
$$
where
$$
F(t_s)=G^5+\eta _1 G^3-{{3\pi } \over 4}G^2+\eta _2G \eqno (4b)
$$
with
$$
G(t_s)\equiv \left[ {{\eta  \over \tau }(t_c-t_s)}\right]^{1/8} \eqno (4c)
$$
$$
\eta _1\equiv {{3,715} \over {8,064}}+{{55} \over {96}}\eta \eqno (4d)
$$
$$
\eta _2\equiv {{9,275,495} \over {14,450,688}}+{{284,875} \over
{258,048}}\eta + {{1,855} \over {2,048}}\eta ^2 \eqno (4e)
$$
where $t_c$ is the time to coalescence from $t_s = 0$, as measured in the
source frame. (Specifying $t_c$ is equivalent to specifying the initial
separation of the orbiting masses.) Taking the time-derivative of Eq.\ 4a
gives the orbital angular frequency $\omega _s \equiv d\phi /dt_s$ in the
source frame. The orbital phase is thus a complicated function of $t_s$,
parameterized by the four parameters $\{\phi_{0},\eta,\tau,t_c\}$.

The gravitational wave detectors are assumed to be of two types: an
ecliptic plane detector, like OMEGA [3], and a precessing-plane detector,
like LISA [4].  Each of these detectors consists of three coplanar
laser-beam arms that form an equilateral triangle with spacecraft at each
vertex of the triangle.  The three arms define the detector plane. An
interferometer is created by combining the beams in adjacent arms, the
three possible combinations representing only two independent information
streams.  The signal received in the detectors is a phase measurement of
the laser interferometer and may be written
$$
{{\phi (t)} \over {\nu T}}=h(t)=F_+(t)h_+(t)+F_\times (t)h_\times (t)
\eqno (5)$$ where $\nu$ is the laser frequency and $T$ is the
light-travel-time along the interferometer arm. The time-dependent
functions $F_+(t)$ and $F_\times (t)$ are the beam-pattern functions for
the interferometer pair. For both the precessing-plane case and the
ecliptic plane case, the arms in a given interferometer pair make a
$60^{\circ}$ angle with respect to each other, and the beam-pattern
functions for the interferometer formed by one such pair are
$$\eqalign{
F_+ = &{\textstyle{{\sqrt 3} \over 2}}\left[ {{\textstyle{1 \over 2}}(1+\cos
^2\theta _D)\cos 2\phi _D\cos 2\psi _D }\right. \cr &\left. {-\cos \theta
_D\sin 2\phi _D\sin 2\psi _D } \right]} \eqno (6a)$$
$$\eqalign{
F_\times  = & {\textstyle{{\sqrt 3} \over 2}}\left[ {{\textstyle{1 \over
2}}(1+\cos ^2\theta _D)\cos 2\phi _D\sin 2\psi _D }\right. \cr &\left.
{+\cos\theta _D\sin 2\phi _D\cos 2\psi _D } \right]} \eqno (6b)$$ where
$\theta _D$ and $\phi _D$ are the instantaneous angular coordinates of the
source measured relative to the frame of the detector and $\psi _D$
specifies the orientation of the binary's principal polarization axes
around the fixed line of sight. The latter variable can be understood as
follows. If the orbital inclination $i$ to the line of sight is not zero,
the quasicircular binary orbit will look elliptical to a viewer at the
detector. The angle $\psi _D$ specifies the orientation of the major axis
of the ellipse as viewed by this observer, measured in the plane
perpendicular to the line of sight and from a reference direction that is
perpendicular to both the line of sight and the normal to the plane of the
detector array. If $\hat{L}$, $\hat{n}$, and $\hat{z} _D$ are unit vectors
parallel to the binary's conserved angular momentum, the direction of the
line of sight, and the normal to the detector array, respectively, then we
may define $\psi_D$ via
$$
\tan \psi _D = {{\hat L\cdot [\hat n\times (\hat z_D\times \hat n)]} \over
{\hat L\cdot (\hat z_D\times \hat n)}} \eqno (7)$$

In both the precessing-plane and ecliptic-plane case, the equilateral triangle
arrangement means that the detector array possesses two independent pairs of
interferometer arms, so we will have two independent signals $h_k(t)$ (where
$k$ = 1, 2).  Because one pair of arms is rotated $60^{\circ}$ relative to the
other in the plane of the detector array, the value of $\phi _D$ will depend
on the choice of interferometer pair as well, and so is denoted $\phi _{D,k}$.

For the ecliptic plane case, the zenith of the detector plane is the same as
the zenith of the ecliptic, so $\theta _D = \Theta$ and $\psi _D = \psi$. If
the satellites orbit the Earth with angular frequency $\omega _d$, then the
apparent azimuth of the source relative to the detector arm will be given by
$$
\phi _{D,k}=\alpha _k(t) \quad\quad{\rm where}\quad\quad \alpha
_k(t)\equiv\Phi -\omega _dt+\alpha _{0k}
$$
Here the constant $\alpha _{0k}$ specifies the orientation of the
interferometer pair at $t$ = 0. While we must have $\alpha _{02} = \alpha
_{01}+\pi/3$, the constant $\alpha _{01} \equiv \alpha_0$ is arbitrary. In
terms of these variables, the beam-pattern functions become
$$
\eqalign{ F_{+,k}  = & {\textstyle{{\sqrt 3} \over 2}}\left[ {{\textstyle{1
\over 2}}(1+\cos ^2\Theta )\cos 2\alpha _k\cos 2\psi}\right.  \cr &\left.
{-\cos \Theta \sin 2\alpha _k\sin 2\psi } \right]} \eqno (8a)$$
$$\eqalign{
F_{\times ,k}  = & {\textstyle{{\sqrt 3} \over 2}}\left[ {{\textstyle{1 \over
2}}(1+\cos ^2\Theta )\cos 2\alpha _k\sin 2\psi}\right.  \cr &\left.
{+\cos\Theta\sin 2\alpha _k\cos 2\psi } \right]} \eqno (8b)$$

In the precessing-plane case, the orientation of the detector plane changes
with time, making the expressions for the beam pattern functions more
complicated.  In this case, we have
$$
\cos \theta _D = {\textstyle{1 \over 2}}\cos \Theta
 - {\textstyle{{\sqrt 3} \over 2}}\sin \Theta \cos \beta
\eqno (9a)$$
$$
\phi _{D,k} = \alpha _k(t) +\tan ^{-1}\left[ {{{\sqrt 3\cos \Theta +\sin
\Theta \cos \beta } \over {2\sin \Theta \sin \beta }}} \right] \eqno (9b)$$
where $\alpha _k(t) \equiv \Omega t + \alpha _0k$ and $\beta (t)$ specifies
the angular position of the detector array's center of mass in the plane of
the ecliptic. Taking account of the way that we have defined $\Phi$ = 0, the
quantity $\beta (t)$ is given by:
$$
\beta (t)\equiv \Omega t-\Phi . \eqno (9c)$$ One can also show that $\psi _D$
in this case is given by
$$
\tan \psi _D= {{-a\cos \psi +b\sin \psi } \over {a\sin \psi +b\cos \psi }}
\eqno (9d)
$$
where $a \equiv \sqrt 3\sin \beta$ and $b \equiv \sqrt 3\cos\Theta \cos
\beta + \sin \Theta$.

In the sections that follow, these expressions will be used to model the
detected signal from a coalescing binary. The ability of the signal to
determine the astrophysical parameters of the source will be simulated via
a linear least-squares estimation process in which partial derivatives of
the received signal with respect to the parameters are accumulated over a
year of time-domain data and an information matrix is formed, weighted by
the inverse of the {\it rms} noise in the detector.  The inverse of the
information matrix is the covariance matrix, whose diagonal terms give the
uncertainty to be expected in each parameter.  Details of the procedure we
use are given in MH and will not be repeated here.  There is, however, one
important difference between the covariance study in MH and the method we
use here. This is in the way that the higher harmonics of the waveform are
handled in the analysis.

The {\it rms} noise that was assumed in the covariance studies in MH was
calculated from the published noise spectra for each mission (OMEGA and
LISA) using
$$\sigma_y=\sqrt{S_n(f)\Delta f} , \eqno (10) $$
where $\sigma_y$ is the rms noise in the detector, $S_n(f)$ is the
published noise spectral density at the frequency $f$, taken to be the
fundamental frequency of the source, and $\Delta f=1/(2\Delta t)$ is the
bandwidth, equal to the Nyquist frequency with sample time $\Delta t$.
The Appendix in MH explains that the noise in the space detectors is
expected to be red noise with a power spectrum that falls off roughly as
$f^{-4}$, and then justifies the use of Eq.\ 10, even in the presence of
red noise, for the case where the signal is dominated by a single
fundamental frequency.  However, when a signal contains higher harmonics
of the fundamental frequency, the analysis of red noise becomes more
complicated.

Let us consider a time series $y(t)=h(t)+n(t)$, where $h(t)$ is the signal and
$n(t)$ is the red noise.  To avoid bias in the parameter estimation, the time
series of length $T$ is first ``prewhitened'' by passing it through a linear
filter, giving
$$x(t)=F(t)*y(t)\equiv\int_0^TF(t-\tau)y(\tau)d\tau . \eqno(11) $$
The effect of the filter on the signal gives $g(t)=F(t)*h(t)$ and the effect
on the noise is $m(t)=F(t)*n(t)$.  The filter $F(t)$ is chosen so that the
power spectrum of $m(t)$ will be flat.  The Fourier transform of the filtered
signal and noise are simple products, $g(f)=F(f)h(f)$ and $m(f)=F(f)n(f)$,
where $F(f)$ is the transfer function of the filter.  When the noise spectrum
$S_n(f)=n^2(f)$ has a $f^{-4}$ power-law behavior, the prewhitening filter is
$F(t)=d^2/dt^2$, with  transfer function $F(f)=4\pi^2f^2$.

The signal-to-noise ratio for the filtered data is given by $({\rm
SNR})^2=\langle g^2(t)\rangle/\langle m^2(t)\rangle$, where the angle brackets
denote a time average.  The mean squared signal and noise strengths may in
turn be written in terms of their spectral densities as
$$\eqalign{({\rm SNR})^2&={\int_{f_L}^{f_H}\, S_g(f)df\over
                         \int_{f_L}^{f_H}\, S_m(f)df}\cr
                        &= {\int_{f_L}^{f_H}\, F^2(f)S_h(f)df\over
                          S_m(f_H-f_L)}}    \eqno (12) ,$$
where, in the last step, the fact that $S_m(f)={\rm const}$ has been used to
complete the integral in the denominator and the fact that $g(f)=F(f)h(f)$ has
been used to expand the numerator.

In MH, the signal $h(t)$ was assumed to be monochromatic, at frequency
$f_0$, so that the power spectrum of $h$ would be
$S_h(f)=\delta(f-f_0)S_h(f_0)$. However, when many harmonics are present,
the power spectrum of the signal will be a series of delta functions, one
at each of the harmonics.  Thus we will have
$S_h(f)=\Sigma_i[\delta(f-f_i)S_h(f_i)]$, which will complicate the
numerator of Eq.\ 12.  In the denomonator, because $S_m(f)$ is constant,
its relation to $S_n(f)$ may be worked out at any frequency desired; we
choose $f_0$.  Eq.\ 12 then becomes
$$\eqalign{({\rm SNR})^2
&={\Sigma_i F^2(f_i)\int_{f_0}^{f_H}\, \delta(f-f_i)S_h(f_i)df\over
    F^2(f_0)S_n(f_0)(f_h-f_0)}\cr
&=[S_n(f_0)(f_H-f_0)]^{-1}\Sigma_i{F^2(f_i)\over F^2(f_0)}\langle
h_i^2(t)\rangle} \eqno(13) $$ where we have assumed that the data have
been high-pass filtered with cutoff at $f_L=f_0$.  In the the last step in
Eq.\ 13, we have recognized the power in each harmonic $\langle
h_i^2(t)\rangle$ as the integral over the appropriate spike of the
spectral density.  It should be remembered that the $h_i(t)$ are the basic
signals in the detector, before prewhitening.

The weighted information matrix used in the covariance analysis in this
paper is found from Eq.\ 13. To calculate the rms noise, we take the noise
spectral density at the fundamental frequency and multiply by the
bandwidth $f_H-f_0=1/(2\Delta t)$, the Nyquist frequency $f_H=1/(2\Delta
t)$ being assumed to be much higher than the fundamental frequency $f_0$.
To calculate the effective signal strength, each frequency component of
the $H_\alpha$ in Eq.\ 2 is boosted by the ratio
$F(f_i)/F(f_0)=f_i^2/f_0^2$. The information content of the higher
harmonics will thus be improved over what would be calculated using the
simple waveform.

\vs \n {\bf III.  Position Sensitivity for OMEGA.}

One of the important conclusions of MH was the necessity of including
higher harmonics in evaluating the ability of an ecliptic-plane detector
such as OMEGA to determine the sky position of a coalescing massive black
hole binary.  However, the improved information content of the higher
harmonics, as given by Eq.\ 13, was not noted in that paper.  Here we will
review the conclusions of MH on this question, including the corrected
treatment of the higher harmonic terms.

A covariance study was performed simulating one year of data in
ecliptic-plane and precessing-plane detectors.  The sources were
coalescing massive black hole binaries with a range of masses and with
ecliptic latitudes corresponding to a range in $\Theta$ from 0$\deg$ to
90$\deg$.  The predicted uncertainties in the nine unknown parameters of
the signal $\{\tau,t_c,z,\eta,i,\phi_0,\psi,\Theta,\Phi\}$ were determined
from the covariance matrix of a linear least-squares parameter estimation
process, as discussed in MH.  The calculated uncertainties in $\Theta$ and
$\Phi$ are combined to give a solid angle uncertainty via
$\delta\Omega=\sin\Theta\,\delta\Theta\,\delta\Phi$.  The results are
shown in Figure 1.  Two sets of curves are shown for each type of detector
(precessing-plane and ecliptic-plane), one (with triangles) representing
the values from MH in which the higher-order harmonics were not properly
boosted and one (with squares) representing the new results with the
correct treatment of higher harmonics included.

As may be seen in the figures, the boosted harmonics help in the
determination of the angular position of the source in the sky for the
ecliptic-plane configuration, especially at the middle ecliptic latitudes.
By contrast, the precessing-plane configuration is little affected by the
higher harmonics, except, surprisingly, in the case of two $10^5$\Msun
coalescing black holes.  The reason for this anomaly is that, at the
lowest frequencies (largest black hole binaries) the position
determination is supplied by the modulation of the signal created by the
precessing detector plane, while, at the highest frequencies (smallest
black hole binaries) the determination is dominated by the Doppler
modulation provided by the motion of the detector around the sun.  At the
middle frequency, near $10^{-4}$ Hz, neither effect is able to provide
strong position information independent of the binary orbit inclination
$i$, and it is the higher harmonic of the gravitational wave waveform,
even for the precessing-plane case, that allows the position to be found.
In MH, we demonstrated the value of the higher-order harmonics for the
ecliptic-plane case by plotting the $\Omega$ uncertainties with and
without higher harmonics included (reference [2], Fig. 6). However, we did
not investigate all black hole masses for the precessing-plane case and
incorrectly concluded that, "artificially suppressing the higher-order
terms in the waveform does not change the angular uncertainties very
much."  The true situation is shown in Fig. 2, where the uncertainty in
$\Omega$ is plotted versus $\Theta$ for two $10^5$\Msun black holes and
two $10^6$\Msun holes, with and without the higher-order harmonics
included.  The harmonics clearly contribute nothing for $10^6$\Msun holes,
but make a substantial contribution in the case of $10^5$\Msun holes.

As was pointed out in MH, the information content of the higher harmonics
may be understood in the following way.  We may consider the signals seen
at each of the two detectors as depending on three effects. The first
effect is the monotonic increase in the frequency of the source, as given
by Eqs. 4. If this increase is written by expanding the frequency in a
Taylor series, then the behavior can be expressed in terms of the
derivatives $\omega_0$, $\dot\omega_0$, $\ddot\omega_0$, etc. As seen in
Eqs. 4, these derivatives are linked to the basic variables $\eta$,
$\tau$, and $t_c$. Observation of the time series will determine the
frequency derivatives and will thus give $\eta$, $\tau$, and $t_c$
independently of any other features of the observed signal. The second
effect is the variation of the signal with orientation of the detector, as
given by Eqs. 6. However, the $F_{+,\times}$, which depend explicitly on
$\Theta$, $\Phi$ (through $\alpha_k$), and $\psi$, are not seen directly
in the signal, but only in convolution with the third effect, the
amplitudes of the two polarizations of the waves, as given by Eq.\ 1. The
$h_{+,\times}$ are determined by the already-known $\tau$, $\eta$, and
$\omega$, and by the unknown parameters $R_L$, $i$, and $\phi_0$. There
are thus six unknown parameters ($\Theta$, $\Phi$, $\psi$, $R_L$, $i$ and
$\phi_0$) that must be determined from the waveform, without any help from
the frequency derivatives. If only the fundamental frequency of the
gravitational wave were present (Eqs. 2a--b), then each detector would see
only a single harmonic, whose amplitude and phase would be the only
observables. For two detectors, there would be two amplitude observables
and two phase observables, but this would not be enough to determine the
six unknown parameters, $\Theta$, $\Phi$, $\psi$, $R_L$, $i$, and
$\phi_0$. However, if a second harmonic of the wave is included (the
harmonics $H^{(1/2)}$ of Eqs. 2c--d), then a Fourier analysis of the
detected signals will determine amplitudes and phases for both harmonics.
Since the mix of phases and amplitudes between the two harmonics depends
on $i$ and $\phi_0$, there will be nontrivial information in these
additional terms, and all six gravitational wave parameters can be
determined from the eight observed quantities. The ability of the higher
harmonics to add information, however, is quenched as $\Theta\rightarrow
0$, because the form factors depend on $\Theta$ only through $\cos \Theta$
and also because $\Phi$ and $\psi$ become degenerate near the pole of the
ecliptic.

\vs

\n {\bf IV.  Masses of chirping binaries}

\nobreak It has long been known that, for a purely monochromatic signal
from a binary star, the total mass of the system and the distance to the
system are completely correlated in the parameterization of the signal,
and so cannot be separately determined via the waveform.  However, when a
binary orbit is tight enough that a measurable change in the orbital
frequency occurs over the time of observation, then a particular
combination of the masses, the so-called ``chirp mass'', may be determined
independently of the distance.  The formula for the chirp mass is derived
from $G(t_s)$ in Eq.\ 4 and is written $${\cal M}=
\left({1-\delta^2\over4}\right)^{{3\over5}}(m_1+m_2)  \eqno (14)$$ Since
the relative mass difference $\delta$ can vary between $-1$ and $1$,
knowledge of ${\cal M}$ gives only a lower bound to the total mass of the
system. It does not determine the total mass (which could in principle
always be infinite, regardless of the value of $\cal M$), and is certainly
not able to determine the two masses separately, at least not without
additional information.

When the higher harmonics of the gravitational wave waveform are included
in the analysis, there is additional information provided.  The
non-linearity that produces the chirp also produces a non-zero higher
harmonic of the gravitational wave, breaking the degeneracy between $\tau$
and $\delta$. The way the information enters is as follows.  The
gravitational wave phase, frequency, and frequency derivative (the chirp)
provide information on the parameters $\{\phi_0,t_c,\tau,\delta\}$ that
define the orbital phase, but it is not possible to determine the four
parameters from the three measured quantities.  (This is in contrast to
the coalescing black hole case, where a frequency double-derivative is
also large enough to be detected).  However, the higher harmonics of Eqs.\
2c and 2d depend on $\delta$ and $i$.  By detecting these harmonics in the
received signal, information on $\delta$ is provided that is independent
of the phase derivatives.

We have performed a covariance study for a chirping neutron star binary
located at the galactic center ($R_L=8\,$Mpc).  Three different initial
frequencies were considered, 200 s, 100 s, and 50 s, and three different
mass ratios were taken, $\delta=0$, $0.1$, and $0.5$.  In each case a set
of inclinations, from $0\deg$ to $90\deg$, were studied.  Other parameters
of the system were chosen arbitrarily.  The results are shown in Figs. 3
and 4. In both figures, the increase of uncertainty at low inclination is
due to the fact that the first harmonic (Eqs. 2c and 2d) is proportional
to $\sin i$.  As $i$ approaches zero, the first harmonic vanishes, leaving
only the much weaker second harmonic as the next term in the gravitational
wave waveform. The effect of the {\it a priori} value of $\delta$ on the
final accuracy of the solution is seen in Fig. 3. In each case $\delta$ is
determined to an uncertainty of about one part in 3. This knowledge breaks
the degeneracy inherent in the chirp mass, allowing the total mass to be
well determined. The increase in uncertainty of the total mass with
increasing $\delta$ is a result of a greater correlation between $\tau$
and $\delta$ for larger values of $\delta$. Knowledge of both $\delta$ and
$\tau$ allows the individual masses to be determined to roughly one part
in 3, limited by the uncertainty in $\delta$ . The effect of the choice of
gravitational wave frequency on the uncertainties is shown in Fig. 4,
where, at higher frequencies, both the frequency derivative and the value
of $\varepsilon$ (Eq.\ 1) are larger, allowing both $\tau$ and $\delta$ to
be determined with better accuracy.

We conclude that gravitational wave analysis of the several neutron star
binaries that are expected in the galaxy, with frequencies around 0.01\
Hz, will allow important population studies to be made.  Knowledge of the
three-dimensional position parameters will determine space densities and
knowledge of the individual masses will provide the ground truth for
evolutionary models.

\vs

\n {\bf V.  Masses of coalescing supermassive black hole binaries}

A recent paper by Hughes [7] has pointed out a potential problem in
parameter estimation for massive black hole binaries. This is due to the
fact that the more massive binaries ($>10^6{\rm M}_\odot$) will spend very
little time in the frequency band of the detector.  As a result, the slow
modulation of the signal will have little time to produce detectable
effects in the waveform, and several parameters, notably the chirp mass,
the reduced mass, and the luminosity distance to the source, will be very
poorly determined.  Hughes has performed covariance studies showing that,
for a binary of two equal-mass $10^6$\Msun black holes, the determination
of these quantities is marginal at $z=1$ and the solution matrix becomes
singular at $z=3$ (see Hughes [7], Tables 3 -- 7).  However, Hughes's
analysis used only the lowest gravitational wave harmonic, in a manner
similar to that of Cutler and Vecchio.  In conversations, and in his
paper, Hughes acknowledges that the addition of the higher order harmonics
could modify his conclusions.

We have performed covariance studies for parameter recovery for massive
black hole binaries, using the full harmonic analysis.  We consider two
cases in which the parameter estimation was very poor when only the
fundamental quadrupole harmonic was included.  First, we consider the case
of two equal-mass $10^7$\Msun black holes at a redshift $z=3$. In this
case we let the analysis run for a full year, allowing the high LISA noise
at low frequencies to restrict the amount of significant signal available.
Second, we consider two $10^6$\Msun black holes, likewise at $z=3$, and
begin the analysis at a frequency of $10^{-4}\,$Hz\footnote
* {It has been suggested, for engineering reasons, to cut off the
accelerometer control law at $10^{-4}\,$Hz, creating an effective
sensitivity wall at this frequency.}.  In this case, the binary system
provides only about 2.7 days of data before the Post-Newtonian
approximation breaks down ($\sim$10 orbits before coalescence) and we
terminate the simulation.

The output from these two cases is shown in Table 1.  For each case, two
runs are shown, one with the higher-order harmonics of the waveform
included and one where they are set to zero.  The formal uncertainties for
each of the nine parameters are shown in the nine columns.  In both cases
without the higher-order harmonics, the information matrix is degenerate.
When this occurs, the inversion program deletes one of the offending
parameters and a solution of lower rank is found. The asterisks in the
$\delta$ column in both cases without higher-order harmonics included
indicate that the matrix was singular and that this parameter was dropped
from the solution. The relative uncertainty in $\tau$, shown in the second
column, would then be equivalent to the relative uncertainty in the chirp
mass {$\cal M$}.

\ts \sm
\def\m{$\!$}
\line{\hfill \vtop{\tabskip=0pt \offinterlineskip \halign{ \vrule
width.8pt# \tabskip=0pt & \quad # \m\hfil   &\vrule width.8pt# &\hfil \ \
# \m\hfil   &\vrule width.8pt# &\hfil \ \ # \m\hfil   &\vrule width.8pt#
&\hfil \ \ # \m\hfil   &\vrule width.8pt# &\hfil \ \ # \m\hfil &\vrule
width.8pt# &\hfil \ \ # \m\hfil   &\vrule width.8pt# &\hfil \ \ # \m\hfil
&\vrule width.8pt# &\hfil \ \ # \m\hfil   &\vrule width.8pt# &\hfil \ \ #
\m\hfil   &\vrule width.8pt# &\hfil \ \ # \m\hfil   &\vrule width.8pt#
\tabskip=0pt\cr \noalign{\hrule height.8pt}
height5pt&\omit&&\omit&&\omit&&\omit&&\omit&&\omit&&\omit&
 &\omit&&\omit&&\omit&\cr
& {\rm Case 1} && $R_L$ && $\tau$ && $\delta$ && $t_c$ && $i$ & & $\psi$
&& $\phi_0$ && $\Theta$ && $\Phi$ &\cr
height5pt&\omit&&\omit&&\omit&&\omit&&\omit&&\omit&&\omit&
 &\omit&&\omit&&\omit&\cr
\noalign{\hrule height.8pt}
height5pt&\omit&&\omit&&\omit&&\omit&&\omit&&\omit&&\omit&
 &\omit&&\omit&&\omit&\cr
& with && 0.081 && 0.00051 && 0.029 && 0.00018 && 0.15 && 0.26 && 0.27 &&
0.0083 && 0.0093 &\cr
height5pt&\omit&&\omit&&\omit&&\omit&&\omit&&\omit&&\omit&
 &\omit&&\omit&&\omit&\cr
\noalign{\hrule height.8pt}
height5pt&\omit&&\omit&&\omit&&\omit&&\omit&&\omit&&\omit&
 &\omit&&\omit&&\omit&\cr
& without && 0.14 && 0.00049 && *** && 0.00018 && 0.26
 && 0.49 && 0.49 && 0.0081 && 0.0094 &\cr
height5pt&\omit&&\omit&&\omit&&\omit&&\omit&&\omit&&\omit&
 &\omit&&\omit&&\omit&\cr
\noalign{\hrule height.8pt}
height5pt&\omit&&\omit&&\omit&&\omit&&\omit&&\omit&&\omit&
 &\omit&&\omit&&\omit&\cr
& {\rm Case 2} && $R_L$ && $\tau$ && $\delta$ && $t_c$ && $i$ & & $\psi$
&& $\phi_0$ && $\Theta$ && $\Phi$ &\cr
height5pt&\omit&&\omit&&\omit&&\omit&&\omit&&\omit&&\omit&
 &\omit&&\omit&&\omit&\cr
\noalign{\hrule height.8pt}
height5pt&\omit&&\omit&&\omit&&\omit&&\omit&&\omit&&\omit&
 &\omit&&\omit&&\omit&\cr
& with && 0.26 && 0.012 && 0.058 && 0.00015 && 0.43
 && 0.74 && 0.75 && 0.017 &&0.021 &\cr
height5pt&\omit&&\omit&&\omit&&\omit&&\omit&&\omit&&\omit&
 &\omit&&\omit&&\omit&\cr
\noalign{\hrule height.8pt}
height5pt&\omit&&\omit&&\omit&&\omit&&\omit&&\omit&&\omit&
 &\omit&&\omit&&\omit&\cr
& without && 5.0 && 0.011 && *** && 0.00015 && 8.4
 && 16 && 17 && 0.25 && 0.29 &\cr
height5pt&\omit&&\omit&&\omit&&\omit&&\omit&&\omit&&\omit&
 &\omit&&\omit&&\omit&\cr
\noalign{\hrule height.8pt} }} \hfill}

\quote

\n Table 1.  Formal parameter uncertainties for runs with and without
higher-order harmonics included in the simulation.  Case 1 was for two
$10^7$\Msun black holes at $z=3$, coalescing for one year.  Case 2 was for
two $10^6$\Msun black holes, likewise at $z=3$, beginning at an initial
frequency of $10^{-4}\,$Hz, with a resulting data span of only 2.7 days.
The uncertainties for $R_L$ and $\tau$ are relative ({\it i.e.},
$\sigma_\tau/\tau$, {\it etc.}), the uncertainty in $\delta$ is absolute
and dimensionless, the uncertainty in $t_c$ is in years, and the
uncertainties in angular quantities are in radians.

\vs

\rm

While we have only investigated two cases and have neither determined the
average sensitivity over all parameters nor examined the structure of the
sensitivity as a function of the initial parameters, some conclusions can
nevertheless be drawn from Table 1.  First, although the gravitational
wave in Case 1 remained buried in the noise throughout the entire year of
data, the ability of the model to dig into the noise to recover the
parameters of the coalescing massive black hole binary is still
significant, even without the higher-order harmonics (though in this case
one must give up the determination of the individual black hole masses).
It therefore appears a scientifically undesirable thing to give up
low-frequency sensitivity entirely by cutting off the position control
system at too high a frequency.  Second, the importance of including
higher-order terms in the parameter estimation model is clear --- it is
the difference between determining and not determining the individual
masses of the binary system and it significantly improves other
parameters, notably the luminosity distance, $R_L$.

\vs

\n {\bf VI.  Conclusions}

The basic conclusion to be drawn from the last three sections is obvious
--- that it is important to include harmonics of the gravitational wave
waveform when trying to recover source parameters from gravitational wave
data.  An accurate position determination for massive black hole binary
coalescence requires higher-order harmonics for a detector with the
ecliptic-plane configuration and also for a precessing-plane detector at
intermediate frequencies near $10^{-4}\,$Hz.  Higher harmonics, in the
case of a chirping neutron star binary, allow a determination to be made
of the individual masses the stars in the binary and a very accurate
determination to be made of the total mass.  Finally, analysis of the
signals from supermassive black hole binaries need the higher harmonics to
provide the information that is lost due to the short time the systems are
visible in the sensitivity window of the detector.

One of us (RH) would like to acknowledge support from NASA grants
NAG5-11469 and NCC5-579, which enabled this work to be accomplished.

\vfill

\eject

\n {\bf References:} \ts

\font\bo=cmbx10

\n \hangindent=0.2in \hangafter=1 [1] C.\ Cutler and E.E.\ Flanagan, Phys.\
Rev.\ {\bo D49} 2658 (1994); E.\ Poisson, Phy.\ Rev.\ {\bo D54} 5939 (1996);
M.\ Peterseim, O.\ Jennrich, and K.\ Danzmann, Class.\ Quantum Grav.\ {\bo 57}
A279 (1996);  C.\ Cutler, Phys.\ Rev.\ {\bo D57} 7089 (1998); C.\ Cutler and
A.\ Vecchio, {\it Proceedings of the 2$^{nd}$ LISA Symposium (AIP Conference
Proceedings 456)}, 95 (1998); A.\ Vecchio and C.\ Cutler, {\it Proceedings of
the 2$^{\rm nd}$ LISA Symposium (AIP Conference Proceedings 456)}, 101 (1998).

\n \hangindent=0.2in \hangafter=1 [2] T.A.\ Moore and R.W.\ Hellings,
Phys.\ Rev.\ {\bo D65}, 062001 (2001).

\hangindent=0.2in \hangafter=1 \n [3]  R.W.\ Hellings {\it et al.}, {\it
Orbiting Medium Explorer for Gravitational Astrophysics (OMEGA)}, proposal to
NASA Medium Explorer program (1998, unpublished).

\hangindent=0.2in \hangafter=1 \n [4]  P.\ Bender {\it et al.}, {\it LISA
Pre-Phase A Report (second edition)} (1998).

\hangindent=0.2in \hangafter=1 \n [5] L.\ Blanchet, B.R.\ Iyer, C.M.\ Will,
and A.G.\ Wiseman, Class.\ Quant.\ Grav.\ {\bf 13} 575 (1996).

\hangindent=0.2in \hangafter=1 \n [6] A.\ Vecchio and C.\ Cutler, {\it op.
cit.} p. 107.

\hangindent=0.2in \hangafter=1 \n [7] S.A. Hughes, MNRAS {\bf 331} 805-816
(2002).

\vfill

\eject

\n FIGURE CAPTIONS

\n Figure 1.  Angular uncertainty as a function of $\Theta$ for equal-mass
pairs of coalescing supermassive black holes at $z=1$.  In each case,
there are two plots presented for each detector configuration --- one
(triangles) without the higher harmonics boosted according to Eq. 13 and
one (squares) where the harmonics have been properly boosted.  The total
mass of the system for each case is displayed in the box on each plot.

\n Figure 2.  Angular uncertainty as a function of $\Theta$ for the
precessing-plane configuration (LISA) for the $2\times10^5$\Msun case of
Fig. 1b and the $2\times10^6$\Msun case of Fig. 1c.  In each case, one
plot is shown where the higher harmonics have been completely suppressed
({\it without}) and one where they are included, boosted in the proper way
({\it with}).

\n Figure 3.  Uncertainties for $\delta=(m_1-m_2)/(m_1+m_2)$ (solid line)
and relative uncertainties for $M_{\rm tot}=m_1+m_2$ (dotted line), as
functions of the source inclination $i$ to the line of sight, for a
neutron star binary with total mass $M_{\rm tot}=2.8$\Msun and initial
frequency $f=0.01\,$Hz. The three curves for each parameter are for three
{\it a priori} values of $\delta$.

\n Figure 4.  Uncertainties for $\delta=(m_1-m_2)/(m_1+m_2)$ (solid line)
and relative uncertainties for $M_{\rm tot}=m_1+m_2$ (dotted line), as
functions of the source inclination $i$ to the line of sight, for a
neutron star binary with total mass $M_{\rm tot}=2.8$\Msun and with
$\delta=0$. The three curves for each parameter are for three initial
gravitational wave periods.

 
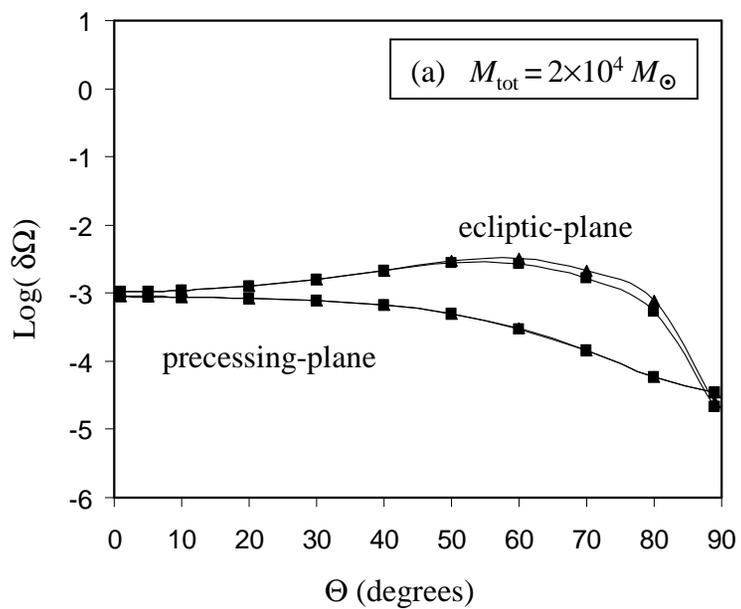
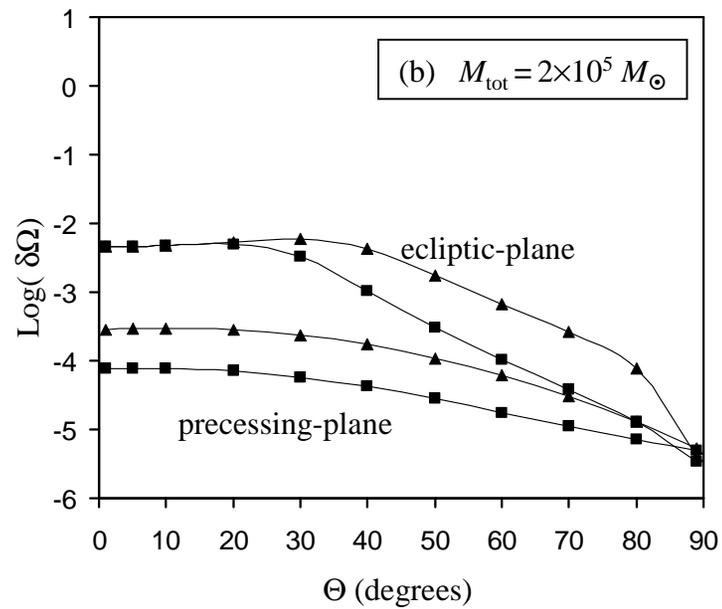
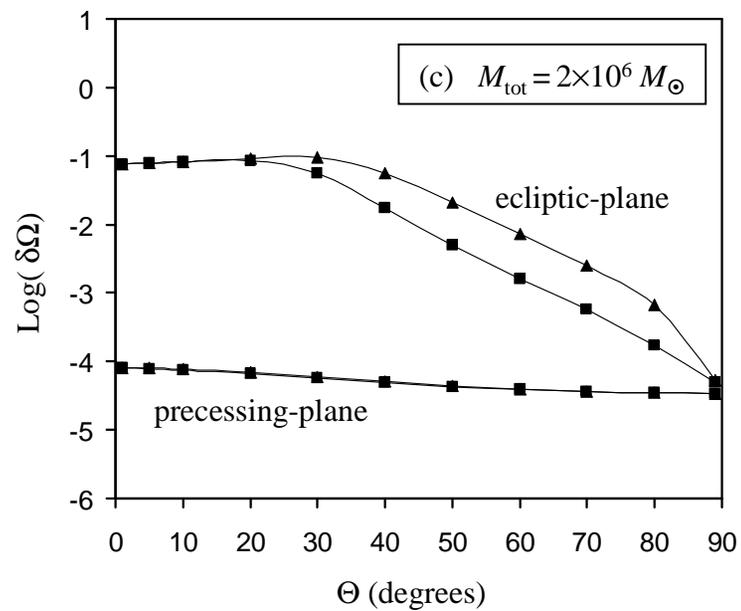
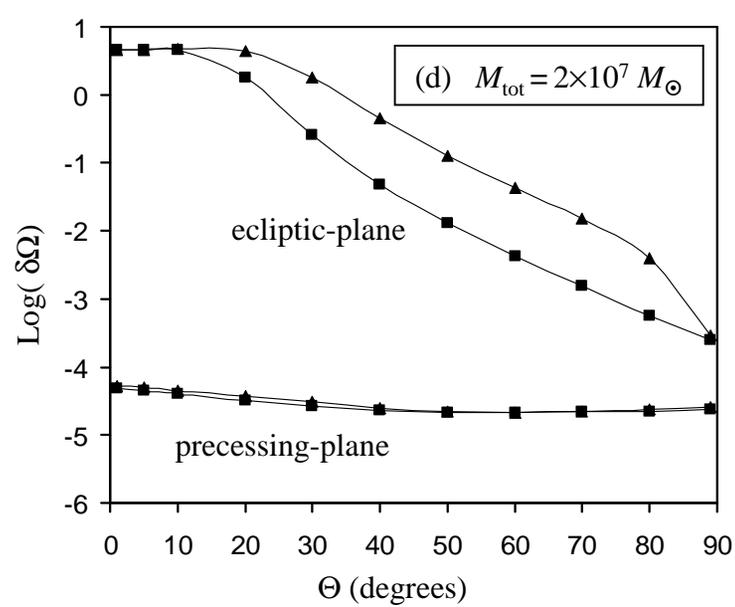
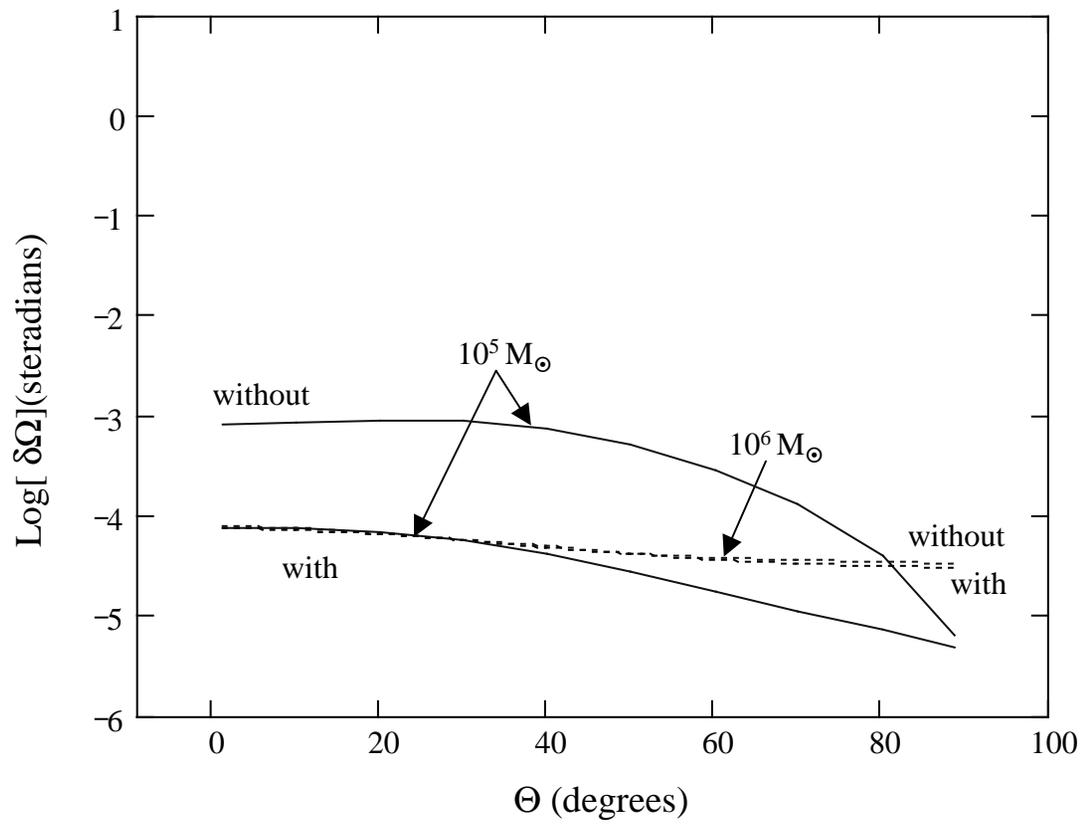
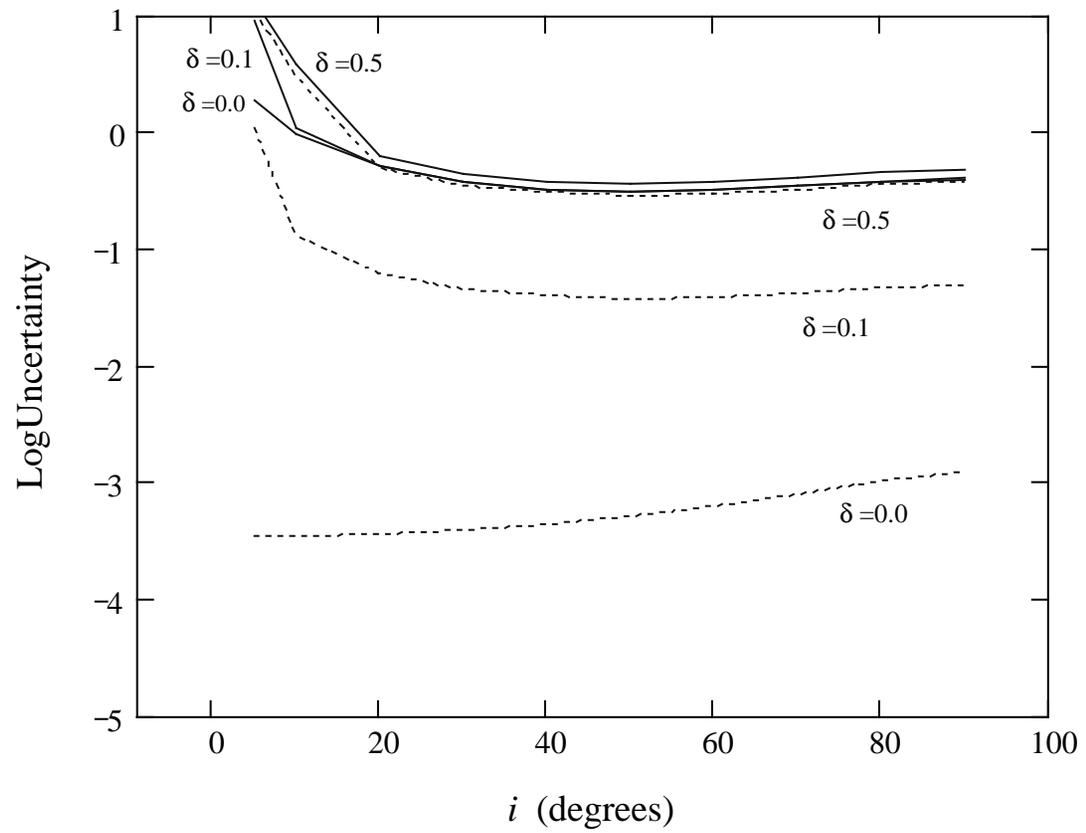
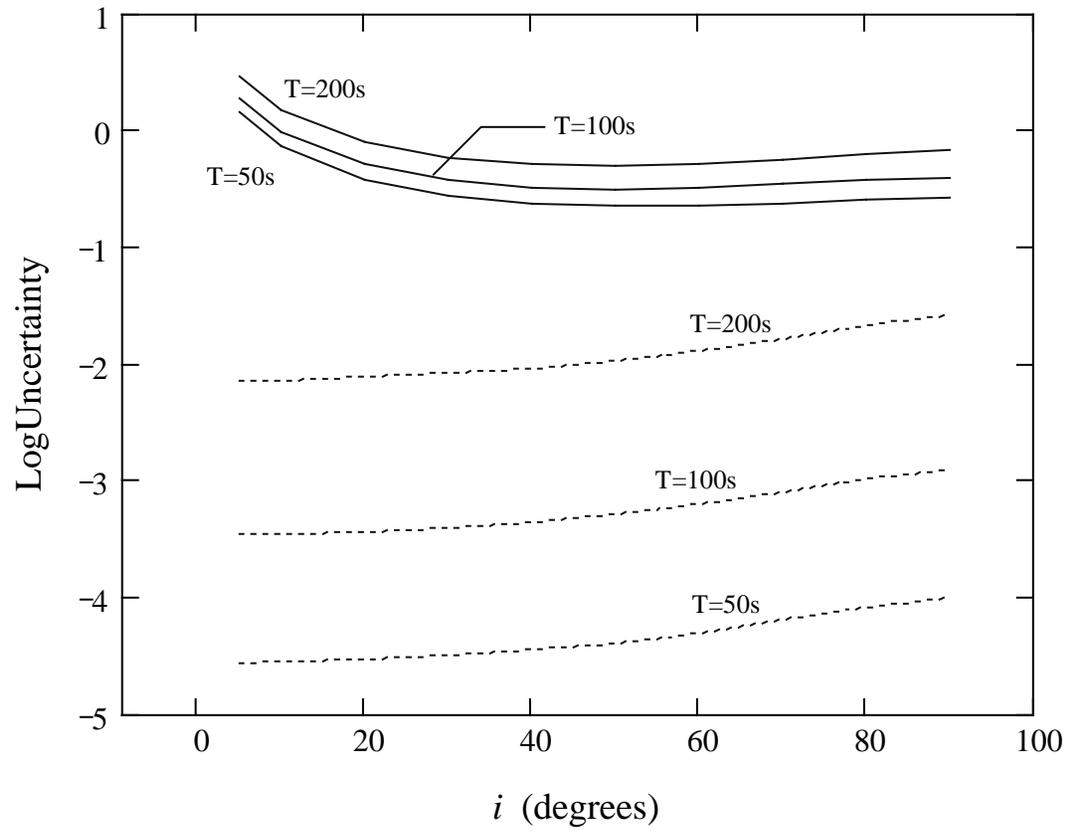
\bye